\documentclass[doublecol,linenumbers]{epl2} % for 2 columns style with line numbers
\title{Room temperature superconductivity dome at a Fano resonance in superlattices of wires}
\author{Maria Vittoria Mazziotti\inst{1,2}  \and Thomas Jarlborg\inst{1,3}  \and  Antonio Bianconi\inst{1,4,5}   \and  Antonio Valletta\inst{6}}
\shortauthor{M. V. Mazziotti \etal}

\institute{                    
\inst{1} RICMASS, Rome International Center for Materials Science Superstripes, via dei Sabelli 119A, 00185 Rome, Italy\\
\inst{2} Department of Mathematics and Physics, Roma TRE University, via della Vasca Navale 84, 00146 Roma, Italy\\
\inst{3} DQMP, University of Geneva, 24 Quai Ernest-Ansermet, CH-1211 Geneva 4, Switzerland\\
 \inst{4} Institute of Crystallography, CNR, via Salaria Km 29.3, 00016 Monterotondo, (Roma) Italy\\
 \inst{5} National Research Nuclear University MEPhI (Moscow Engineering Physics Institute), 115409 Moscow, Russia\\
 \inst{6} Institute for Microelectronics and Microsystems IMM,CNR, via del Fosso del Cavaliere, 100, 00133 Roma, Italy\\  }

\abstract{Recently room temperature superconductivity with $T_C$=$15$ degrees Celsius 
has been discovered in a pressurized complex ternary hydride, $CSH_x$ which is
a carbon doped $H_3S$ alloy. The nanoscale structure of $H_3S$ is a particular realization of the 1993 patent claim of  superlattice of quantum wires for room temperature superconductors where the maximum $T_C$ occurs at the top of a superconducting dome. Here we focus on the electronic structure of materials showing nanoscale heterostructures at atomic limit made of a superlattice of quantum wires like hole doped cuprate perovskites, organics, $A15$ intermetallics and pressurized hydrides.  We provide a perspective of the theory of room temperature multigap superconductivity in heterogeneous materials tuned at a Fano Feshbach resonance (called also shape resonance) in the superconducting gaps focusing on $H_3S$ where the maximum $T_C$ occurs where the pressure tunes the chemical pressure near a topological Lifshitz transition. Here the superconductivity $dome$ of $T_C$ versus pressure is driven by both electron-phonon coupling and contact exchange interaction. We show that the $T_C$  amplification up to  room temperature is driven by the Fano Feshbach resonance between a superconducting gap in the anti-adiabatic regime and other gaps in the adiabatic regime. In these cases the $T_C$ amplification via contact exchange interaction is the missing term in conventional multiband BCS and anisotropic Migdal-Eliashberg theories including only Cooper pairing}
\begin{document}

\maketitle

\section{Introduction}

\begin{figure}
	\centering
	\includegraphics[scale=0.95]{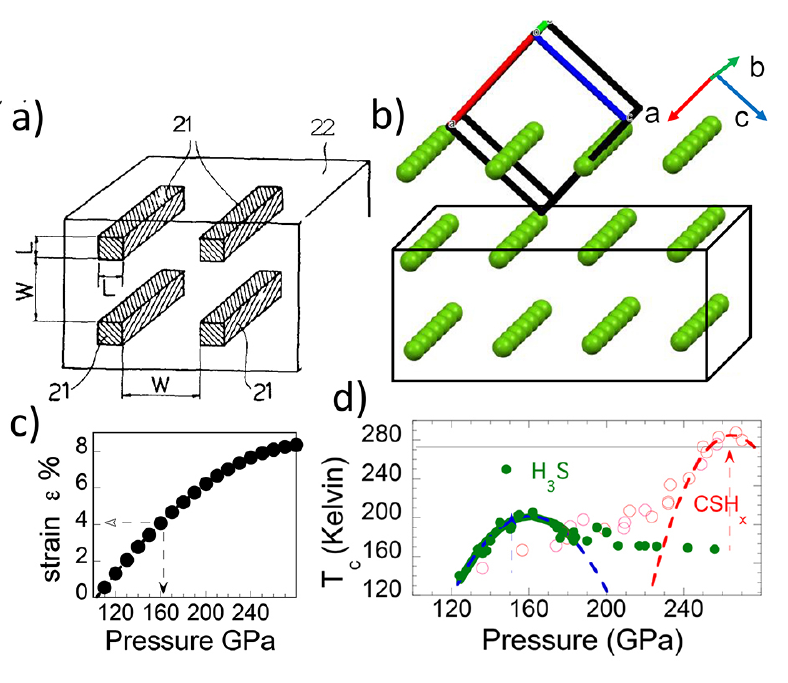}
\caption{
\textit{Panel} \textbf{a}: The heterostructure of superconducting wires (21) embedded in the spacer material (22) presented in the claim of the patents \cite{bianconi1998high,bianconi2001process} for realization of  room temperature superconductors made of superlattices of quantum wires. 
 \textit{Panel} \textbf{b}: The nanoscale structure of $H_3S$ \cite{goncharov2017stable,duan2017structure} with the $Im\bar{3}m$ symmetry formed by two sublattices$:$ the hydrogen wires (green dots) and the $SH$ spacers. The figure shows only the sublattice of hydrogen wires in the (101) planes embedded in the $SH$ spacer material.
\textit{Panel}  \textbf{c} shows
 the compressive strain of the lattice as a function of pressure from the structural transition at $103\ GPa$. The compressive strain at $160\ GPa$  is $4$ percent.
 \textit{Panel} \textbf{d} shows the critical temperature of $H_3S$ and $CSH_x$ hydrides 
as a function of pressure \cite{droz,duan2017structure,100}.  
  }
\label{fig.1}     
\end{figure}

Superconductivity was discovered in 1911 in a pure elemental metal at a temperature close to zero Kelvin and it was explained by pairing mediated by electron-phonon interaction \cite{bcs,migdal1958interaction,mcm}. Today, after 110 years, a ternary hydride $CSH_x$ \cite{100} at $P=265\ GPa$ with the superconducting critical temperature  $T_C$=$15$ degrees Celsius has been discovered.
$CSH_x$  alloy is the outcome of the material research focusing on chemical doping \cite{Sakanishi2018first, durajski2018gradual, kruglov2017refined} of pure $H_3S$ \cite{duan,droz}. While the structure of $H_3S$ and its variation with pressure is well established \cite{goncharov2017stable, duan2017structure}, the structure of $CSH_x$ is still object of research  \cite{bykova2020structure}.
It has been shown  \cite{bia15a,jar16} that $H_3S$ is made of a 3D heterostructure of one-dimensional (1D) hydrogen wires shown in 
Fig. \ref{fig.1}. 
This nanoscale structure is a practical realization of the heterostructures at atomic limit formed by a superlattice of quantum wires described in Fig. {\ref{fig:4.5}} of the claims in the 1993 patent shown in Fig. \ref{fig.1} \cite{bianconi1998high,bianconi2001process} for room temperature superconductors (RTS) 
driven by shape resonances \cite{bianconi2013shape,bianconi1994instability,bianconi1996determination,2}. 
In this proposal for the mechanism of RTS both the internal pressure, due to lattice misfit between nanoscale modules, and the external pressure are used to tune the chemical potential in the proximity of Lifshitz transitions for strongly interacting electrons \cite{08,09}.
In $H_3S$ the external pressure tunes the compressive strain of short H-H bonds in the wires, and the long bonds between wires up to reach the optimal strain (shown in Fig. \ref{fig.1}) for the maximum of $T_C$ at the top of a superconductivity $dome$,  
Ashcroft \cite{ash} suggested to search for room temperature superconductors by applying external pressure to composite hydrides made of hydrogen nanoscale modules where the bond lengths are already  compressed by internal chemical pressure due to misfit strain in a nanoscale heterostructure. 
 
\section{Hole doped cuprate perovskites}

\begin{figure}
	\centering
\includegraphics[scale=0.8]{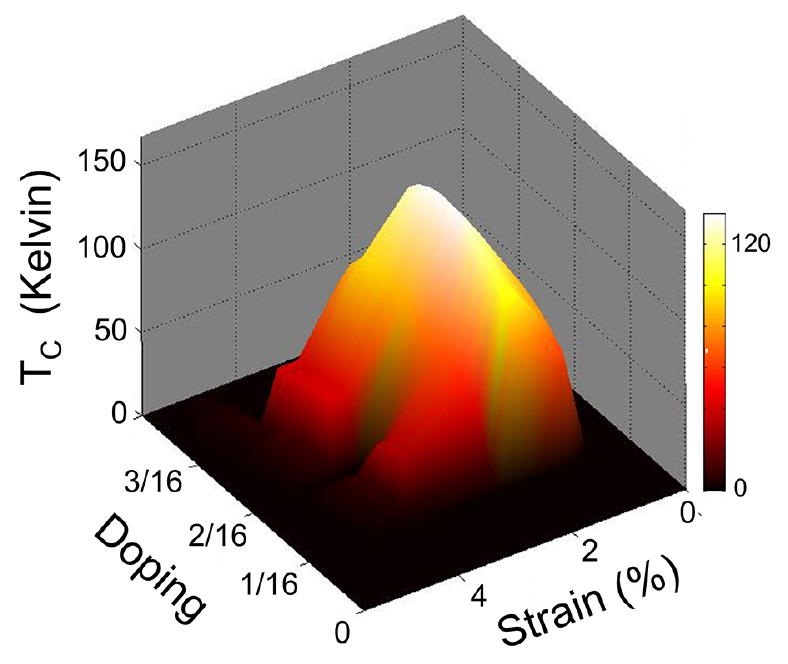}
\caption{
The superconducting $dome$ for all cuprate hole doped superconductors belonging to many cuprate families. The critical temperature $T_C$  is plotted as a function of $doping$, the number of holes for Cu site, and $strain$, $\epsilon=100(R_0-R)/R_0$, where R is the average $<CuO>$ bond length and $R_0=197\ pm$, due to misfit between $CuO_2$ layers and rock salt oxide layers. $T_C=130 K$ at the top of the $dome$ is reached at the optimum strain, $2$ percent, and doping  $3/16$.
}
\label{fig:4.2}     
\end{figure}

The correlated disorder in quantum complex matter has  been unveiled in these last two decades
 by novel imaging methods using nanoscale X-ray diffraction
\cite{campi2019,fratini2010scale,campi2015inhomogeneity, poccia2011evolution,campi2017nanoscale} 
and XANES or EXAFS methods of X-ray spectroscopy probing local and fast fluctuations  
\cite{della1995atomic, darapaneni2020simulated} showing the formation of 
puddles of superlattices of quantum stripes \cite{bianconi1996determination}.
Nanoscale inhomogeneity is controlled in hole doped cuprate perovskites \cite{2, 3, 4, 5, 6, bianconi2000strain, 8}.  
The critical temperature at ambient pressure is controlled by both misfit strain ($\epsilon$) between nanoscale modules and hole doping  
$\delta$  giving the 3D phase diagram $T_C(\delta,\epsilon)$, which is shown in Fig. \ref{fig:4.2}. 
High pressure shifts the doping values for the maximum critical temperature $T_C=160 K$ \cite{1}
The structure of cuprate perovskites is made of active metallic 2-D $CuO_2$ atomic layers intercalated by insulating spacers.
The band structure for a single flat undistorted $CuO_2$ layer has been calculated assuming only $Cu-O$ $b_{1g}$  orbitals 
in the average crystal lattice, but complexity appears due to non-stoichiometry and nanoscale phase separation in the proximity 
of a electronic topological Lifshitz transition or a metal-insulator transition (MIT) with magnetic or charge-density wave short-range puddles. 
In the proposed superstripes scenario the nanoscale structure is 
made of i) puddles of ordered 1D-chains of oxygen interstitials, 
ii) U stripes of undistorted $CuO$ lattice intercalated by D stripes of distorted lattice, 
iii) incommensurate lattice modulation due to misfit strain iv) charge density waves puddles
 v) spin density wave domains, making the actual electronic structure at nanoscale 
 very different from the band structure of a  homogeneous lattice.
The strained and pressurized hole doped cuprates show complex nanoscale heterostructure made of  nanoscale 1D-chains
 (or stripes) which can be driven toward the maximum of a superconducting $dome$ by pressure and doping.
 Nanoscale phase separation  is a universal feature which appears in  vanadates at a metal to insulator transition \cite{9,10}  
 and near a Lifshitz transition  in diborides \cite{12, campi2006study, agrestini2004substitution, simonelli2009isotope}. 
Phase separation controlled by misfit strain and doping has been observed also in iron based superconductors \cite{13,ricci2010structural}. 
In this work we focus on 3D lattice networks of  
1D chains while the majority of the research efforts in high-$T_C$ superconductivity
was addressed to intrinsic effects like pseudogap, strong electronic correlation, superconducting fluctuations, quantum critical points for a single electronic component in a rigid lattice. 

\section{A15 alloys}

The A15 superconductors, among which $V_3Si$ and $Nb_3Sn$ are most known, have the  $Im\bar{3}m$ lattice symmetry
similar to high pressure phase of metallic $H_3S$.
The semiconductor atoms (group III-IV elements $Si,Sn,Ga,Ge,In, etc$) form a body centred cubic (bcc) lattice in which the metal atoms 
($V,Nb,Cr,Mo..$) form perpendicular one-dimensional chains. 
The DOS at $E_F$ is dominated by the metallic sites, and in the $1970s$ it was much discussed that the electronic
structure was made up by the 1-D chains with very little interchain coupling 
and only minor coupling to the elements in bcc lattice \cite{weger}. 
The DOS would then take an essentially 1-D shape with singular peaks at the van-Hove extremes of the d-bands.
The idea was that the electron filling could put $E_F$ close to the high DOS peaks, where the high $N(E_F)$ 
makes the electron-phonon coupling  $\lambda$ large. This 1-D model gives a plausible explanation for the high $T_C$ 
that was observed among the A15's at the time, i.e. just above $20\ K$. 
The 1-D composition of the electronic structure of such a relatively complicated material as the A15 compound 
seemed very instructive since many physical mechanisms could be clarified without the support of heavy band structure computations.
Later with the development of band structure methods and computers the scientific community accepted the dogma that
the electronic structures of the A15 materials could not be reduced to the 1-D chains \cite{kle78,pic79,jar79,kes78}. 
The results from different methods showed indeed DOS peaks, but also that 
the correct DOS shape depends on band interactions between chains as well as on the
interaction between chains and the bcc positioned atoms. In fact, the correlation between calculations
of electron-phonon coupling and $T_C$ based on the band structure fits well with the experimentally observed $T_C$'s \cite{pap77,arb}. 
Calculations for $V_3Si$ have shown four pieces of Fermi surface with shapes
in agreement with results from angular correlation of positron annihilation \cite{jmp}. 
The interchain and chain-to-bcc-lattice interactions
are not the most important interactions, but it was clear that essential features of the electron structure were missing if not all interactions
were taken into account. 
This is evident from that fact that the DOS of the A15's with metallic elements ($Ir,Pt,Au,..$) on the bcc lattice
is different from the DOS with semiconductor atoms on the bcc positions \cite{jjp83}. 
The atoms on the bcc lattice are sufficiently close to the metal chains, and s-p-orbitals are sufficiently extended to
overlap and hybridize with the d-orbitals on the metal sites and
to destroy the 1-D band that might come from un-hybridized d-states on the chains. 
The other group of A15 superconductors, with heavy 5d metal atoms on the bcc lattice, 
could be believed to have less hybridization with the chains,
since the reach of d-states is shorter than for $s$ and $p$ ones. However, the 5d band is filled and deep below
$E_F$, and the DOS at $E_F$ contains a considerable amount of 6s- and 6p-character. 
The general shape of the DOS is different from the first group
of A15's with $E_F$ moved upwards because of the enhanced valence band filling, but it is not like a 1-D DOS. 
The stability of the A15's with 1-D chains might be due to s- and p-states on the bcc lattice, 
but any 1-D band structure is effectively washed out by hybridization.
So-called canonical band calculations done for the V-d bands only (all 3-d on the 6 V atoms), i.e. no
hybridization with anything else, not even with s or p on the V atoms themselves \cite{jar79}, 
show no pronounced structure for the pure V-d DOS, no sharp peaks, so it did not resemble the true
hybridized d-DOS nor the 1-D DOS model.
This shows that even (d-) hybridization between the chains is sufficient for making the band structure 3-D in the A15's. The importance of pressure and strain for the stability and appearing of charge density waves in A15 compounds was noticed in the $70$-ties 
\cite{01,02}.
Recently new light has been shed by experiments on A15 superconductors and related materials
showing remarkable structural instability \cite{03}, short range charge density waves and nanoscale phase separation \cite{04,05,06,07} near 
 a Lifshitz transition for overlapping bands in presence of electron correlations and small Fermi Surface hot spot where Migdal approximation breaks down \cite{08,09}.
\begin{figure}
	\centering
	\includegraphics[scale=0.6]{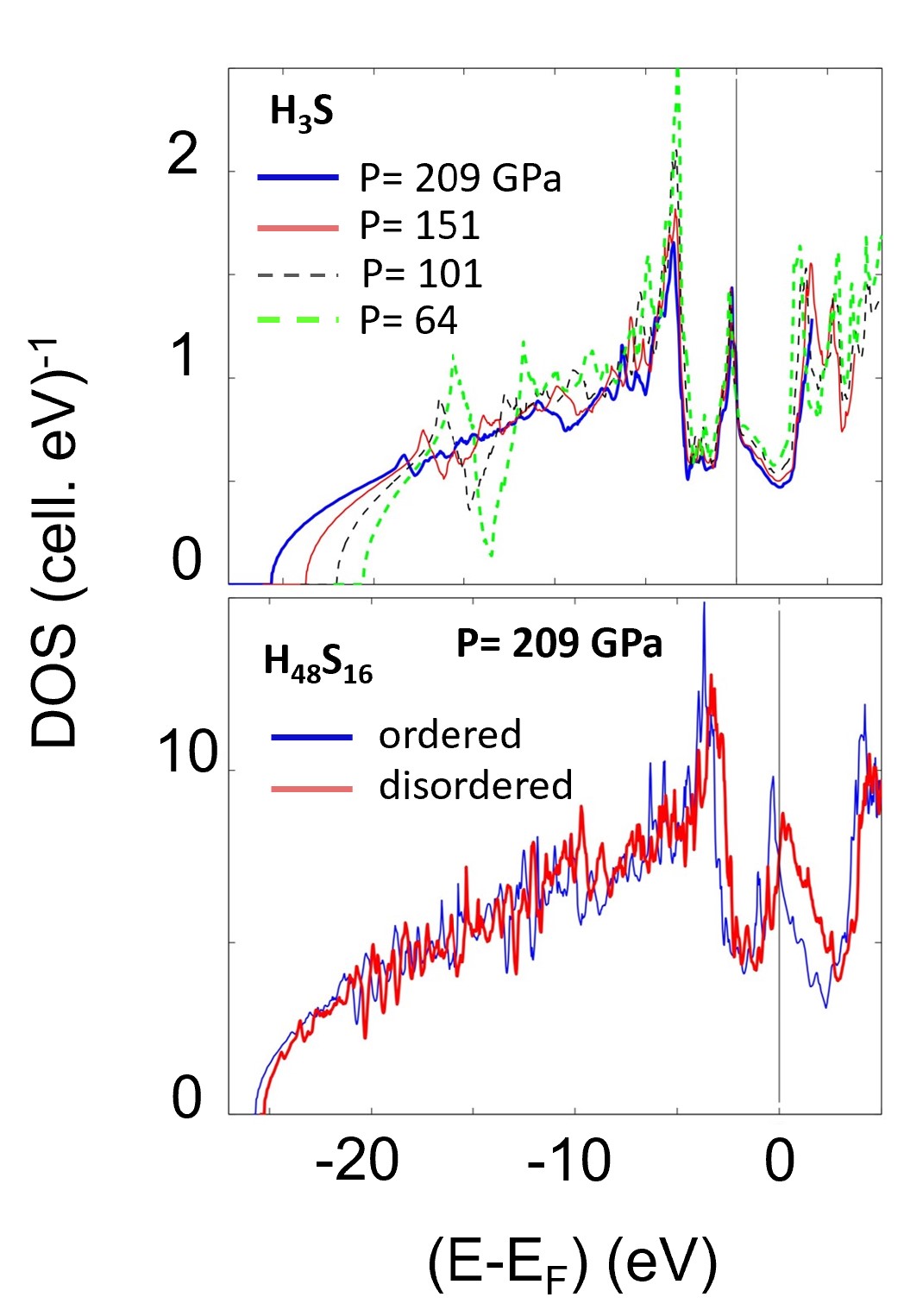}
	\caption{Upper panel: The total DOS for $H_3S$ with variable  lattice constants
		between $5.6$ a.u. and $6.2$ a.u. in the 209 GPa and 64 GPa  pressure range. Lower panel: The DOS for $H_3S$ for a cubic 64-site supercell.
		The (blue) thin line is the DOS for the perfectly ordered cell. The (red) heavy
		line shows the DOS for the disordered cell with $u_S=0.01a$ and $u_H=0.033a$.
		This disorder corresponds to the zero-point motion (ZPM) for S while the disorder for H is less than  ZPM. (From ref. \cite{jar16}) }
	\label{fig2}
\end{figure}
\section{$H_3S$ hydride}

The high pressure phase of $H_3S$ is perhaps not so exotic compared to that of the A15's 
and other high temperature superconductors since the 3-D character remains over a large energy range 
of large portions of the Fermi surfaces, but it shows overlapping bands with quasi 1-D features appear in particular narrow energy regions 
and in particular hot spots of a small Fermi surface near the Fermi level. 
The unusual feature of hydrides is the light hydrogen ion mass. It makes the phonon frequency 
and the vibrational amplitudes of optical modes, involving $H$ ions, very large. 
The popular method for calculation of $\lambda$ and $T_C$ in transition metals 
and their compounds (A15, etc.)
was applied to $H_3S$ before it was synthesized
under pressure ($P$), and despite many uncertainties along 
the computations it correctly predicted a very high $T_C$ \cite{duan}. Other 
calculations, where strong coupling is included through the McMillan parametrization \cite{mcm} of Eliashberg solutions, confirmed these results \cite{papa,bia15a,jar16,bia15}.
Retardation via electron-electron interaction enters through $\mu^*$ in these calculations. 
The single band Eliashberg theory with phonon mediated superconducting mechanism leads to high $T_C$ with high phonon frequency although $\lambda$ is not very large.
The contribution to $\lambda$ from $S$ ions turns out to be comparable to that from $H$,
mainly because of the larger DOS on S. Still, the DOS at $E_F$ per atom is much lower in $H_3S$
than in good transition metal compounds, 
like the A15's, which emphasizes that the high phonon frequencies of hydrogen compounds push $T_C$ to the large values. 
The total band width
is large, and it widens more with $P$, cf Fig. \ref{fig2}. The DOS is therefore expected to diminish with $P$, but charge 
transfers make $E_F$ closer to the peak in the DOS at higher pressure so that $N(E_F)$ is highest for $a$ = $5.6$ $a.u.$ . Orbital overlap and s-to-p-hybridization goes up with pressure, and this is favorable to larger dipole matrix elements in $\lambda$. One might
conclude that the high $T_C$ of $H_3S$ can be understood from this, quite standard, type of calculations. But 
flat bands contribute to the high and narrow DOS peak at $E_F$.
 
The low mass of $H$ also makes the vibrational amplitudes $<u>$ large. At low
temperature the zero-point motion (ZPM) makes $<u> = \frac{3}{2} \hbar \omega / K$, where $K$ is a force constant. 
The amplitudes increase at large $T$ ('thermal disorder') and becomes proportional to $3 k_BT/K$ at high $T$. One can
estimate that $<u>$ approaches $10$ percent of the $H-H$ distance at low $T$ \cite{jar16}. This is much more than in
common transition metal compounds (about $2$ percent), and the electronic structure calculations have to take this into account.
The real lattice is disordered.
 The Madelung potential - a classical variable - is not the same at every H-site, and the
electronic structure - a quantum product - will be different for the disordered lattice compared to that of the perfectly
ordered lattice. This can be seen on physical properties in $fcc$ $Ce$, $FeSi$, purple bronze, etc. \cite{ce,cevib,fesi,bron},
although $<u>$ are much smaller than in $H_3S$. 

In view of the large disorder of $H$ in $H_3S$ one might expect that the DOS-peaks calculated for the ordered lattice will
be washed out by disorder. However, the calculation for supercells of disordered $H_{48}S_{16}$ show no drastic reduction
of the DOS-peaks compared that of the ordered supercell, see Fig. \ref{fig2}. The reason is that a large amount of the DOS at $E_F$ 
comes from $S$ atoms with larger mass and smaller $<u>$. Moreover,  the states on $H$ are s-states, 
which are loosely bound to the nucleus and will not easily follow
the movement of the atom. On the contrary, d- and f-states in transition metals, $Ce$, etc. are tightly bound to the nuclei
and more sensitive to lattice distortions. Another effect from disorder in $H_3S$
is that $E_F$ appears to move relative to the (wider) DOS peak compared to that of the ordered case. 

The band structure of $H_3S$ is very wide, about $2\  Ry$ for the valence bands with  
high band dispersion for most bands below $E_F$ (mainly $s$ and $p$). 
Effects of energy-band broadening in these parts of k-space are less important. 
Flat bands are 
sensitive to disorder, especially at van-Hove singularities. 
Small hole- or electron-pockets (containing   \emph{s-} and  \emph{p-states}  on $H$ and p- and d- on $S$) around these points of the k-space are expected to show fluctuations controlled
by the ZPM. As was discussed above, increased hybridization within the flat bands can compensate
for the wider and less peaked DOS through larger $\lambda$ and $T_C$ are almost the same
for the ordered and disordered supercell. 

The pressure variation of $T_C$ according to the Lifshitz scenario depends sensitively on how the flat bands behave close to $E_F$.
However, a simple scaling of the bands as function of $P$ leads to $T_C$-variations through $3$ main channels; 
$1$: The dipolar matrix element $I$ ($\lambda = \frac{NI^2}{M\omega^2}$) goes up with $P$ because of increasing s-,p-,d- and f-admixture at $E_F$.
$2$: Phonon frequency $\omega$ goes up with $P$ (it is proportional to the square root of the bulk modulus).
$3$: The total DOS $N$ goes down with $P$.
Thus, $T_C$ should go up because of point $1$, and down because of point $3$. Point $2$ gives two opposite effects; the $\omega$-prefactor in the $T_C$-equation has a positive effect as function of $P$, but $\omega^2$ in the denominator of $\lambda$ has a strong negative effect, especially at high
$P$ when the lattice is increasingly stiff. This is seen as a saturation of the calculated $T_C$ at the highest $P$ \cite{jar16}.
 Disorder is real at all $P$, even if it is relatively less important when the lattice becomes very stiff. The widened DOS peak near $E_F$
due to disorder implies an improved stability for $T_C$, and a boost of $T_C$ is possible
 if the position of $E_F$ relative to the peak could be optimized. 
Hence, one should increase $N(E_F)$ through weak
electron doping according to the DOS for disordered lattice, and not by hole doping
as would be suggested from the DOS for perfectly ordered $H_3S$.

\section{The superconducting dome at a Fano resonance}

Following the experimental evidence for the presence of weakly interacting stripes  in hole doped cuprate perovskites \cite{bianconi1994instability,bianconi1996determination} the Bianconi-Perali-Valletta (BPV)  theory was proposed \cite{perali1996, bianconi1997high, bianconi1998superconductivity, valletta1997electronic, perali2012anomalous}.
The BPV theory is based on the solution of the Nikolai Bogoliubov gap equation for a multigap superconductor 
which includes the interference between all pairing channels in the space of configurations following the quantum mechanics formulation of
Gregor Wentzel   \cite{Wentzel}  and Richard Feynman \cite{Feynman}. In this formulation path integrals inspired by the Hamilton principle in classical mechanics and the non locality of the quantum wave-function play a key role. The Wentzel and Feynman approach includes the interference between scattering channels in the space of all configurations which is the outcome of the calculation of the probability amplitude associated with the entire motion of the particles as a function of time, while the Heisenberg and Schroedinger formulation calculates simply the position of the particle at a particular time. 
The path integral formalism of quantum mechanics predicts the configuration interaction between a first closed scattering channel and 
a second open scattering channel which gives a Fano resonance  \cite{Fano} with its characteristic asymmetric line-shape. The Fano line-shape, of the resonance, associated to the negative and positive interference, is the $smoking$ $gun$ which validates the realization of quantum interference. 
These fundamental quantum interference phenomena due to configuration interaction between closed and open scattering channels are called (i) $Fano$  resonance in atomic physics \cite{Fano},  (ii) $shape$ resonance in nuclear physics \cite{Feshbach} and in multigap superconductivity at Lifshitz transitions \cite{perali1996,bianconi1998superconductivity}, $Feshbach$  resonance in ultracold gases \cite{Salomon}.\\
The emergence of these quantum mechanical resonances, called here $Fano$ $resonance$, between different pairing channels is not included in the BCS theory for a single band \cite{bcs} or for overlapping bands \cite{103} as well as in Migdal-Eliashberg \cite{migdal1958interaction} theory for isotropic and anisotropic superconductivity.
 The BPV theory in multigap superconductors predicts the maximum critical temperature by tuning experimentally pressure, strain or charge density  to drive the chemical potential at a Fano resonance between 
i) the pairing scattering channel, called the closed  channel (c), and ii) the pairing scattering channel, called open channel (o).
 The closed  channel (c) forms a Cooper pair in the BEC-BCS crossover regime in the hot spot of the small Fermi surface, appearing at the Lifshitz transition, at the border of the anti-adiabatic regime, where the low Fermi energy $E_{Fc}$ is of the order of the pairing interaction energy 
  ${\omega}_0$. The open channels (o) form Cooper pair in the BCS regime in the large Fermi surfaces with very high Fermi energy $E_{Fo}$ much larger than the pairing interaction energy ${\omega}_0$. The Bogoliubov equation for the gaps is solved together with the density equation which allows the variation of charge superfluid 
 densities in each Fermi surface spot following the opening of different gaps in different Fermi surface spots.  At the Fano resonance the quantum interference effects are controlled by the contact exchange interactions given by
  the overlap of the pair wave-functions i.e.,  the exchange interactions between  pairs in different bands 
  which can be attractive or repulsive like in nuclei where the repulsive exchange interaction gives the Heisenberg force, 
  and the attractive exchange interaction gives the Majorana force.\\
The key idea of the BPV theory is to calculate the exchange contact interaction (neglected in standard BCS theories) between two coexisting condensates in  different Fermi surface spots by the overlap of the wavefunctions. 
Therefore the wavefunctions of electrons are calculated by solving the Schroedinger equation\cite{mazziotti2017possible} or the 
non-relativistic Dirac equation including spin-orbit interactions\cite{mazziotti2021multigaps} within a narrow energy range around the Fermi level of the order of few times the energy cut-off of the electron-phonon pairing interaction.\\
\begin{figure}
	\centering
	\includegraphics[scale=0.65]{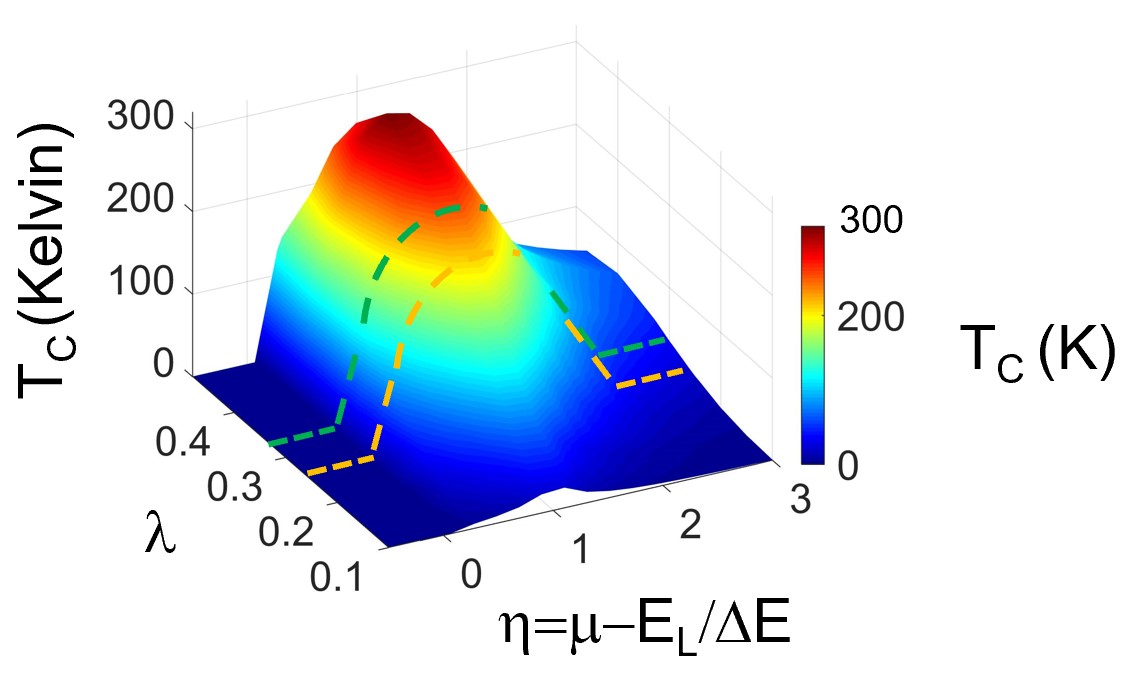}
\caption{The 3D phase diagram of multigap superconductivity in a superconducting superlattice of quantum stripes where  the chemical potential is tuned to a Fano resonance which simulates room-temperature superconductivity in sulphur hydrides superconductors. 
The critical temperature $T_C$ calculated by BPV approach is plotted as a function of $i)$ the electron-phonon coupling $g=g_{33}$ in the upper subband  in the range $0.1<g<0.49$ and $ii)$ the Lifshitz parameter $\eta$ given by the energy separation between the chemical potential  $\mu$  and the position of Lifshitz transition $E_L$, normalized to the transversal energy dispersion $\Delta$E. $\eta$  is tuned in the range $-0.5<\eta <3$.
The yellow (green) dashed line represents the cut of the $dome$ $T_C(\eta)$  for constant $g_{33} = 0.25$ (0.33) coupling.}
\label{fig:4.4}     
\end{figure}
\begin{figure}
\includegraphics[scale=0.4]{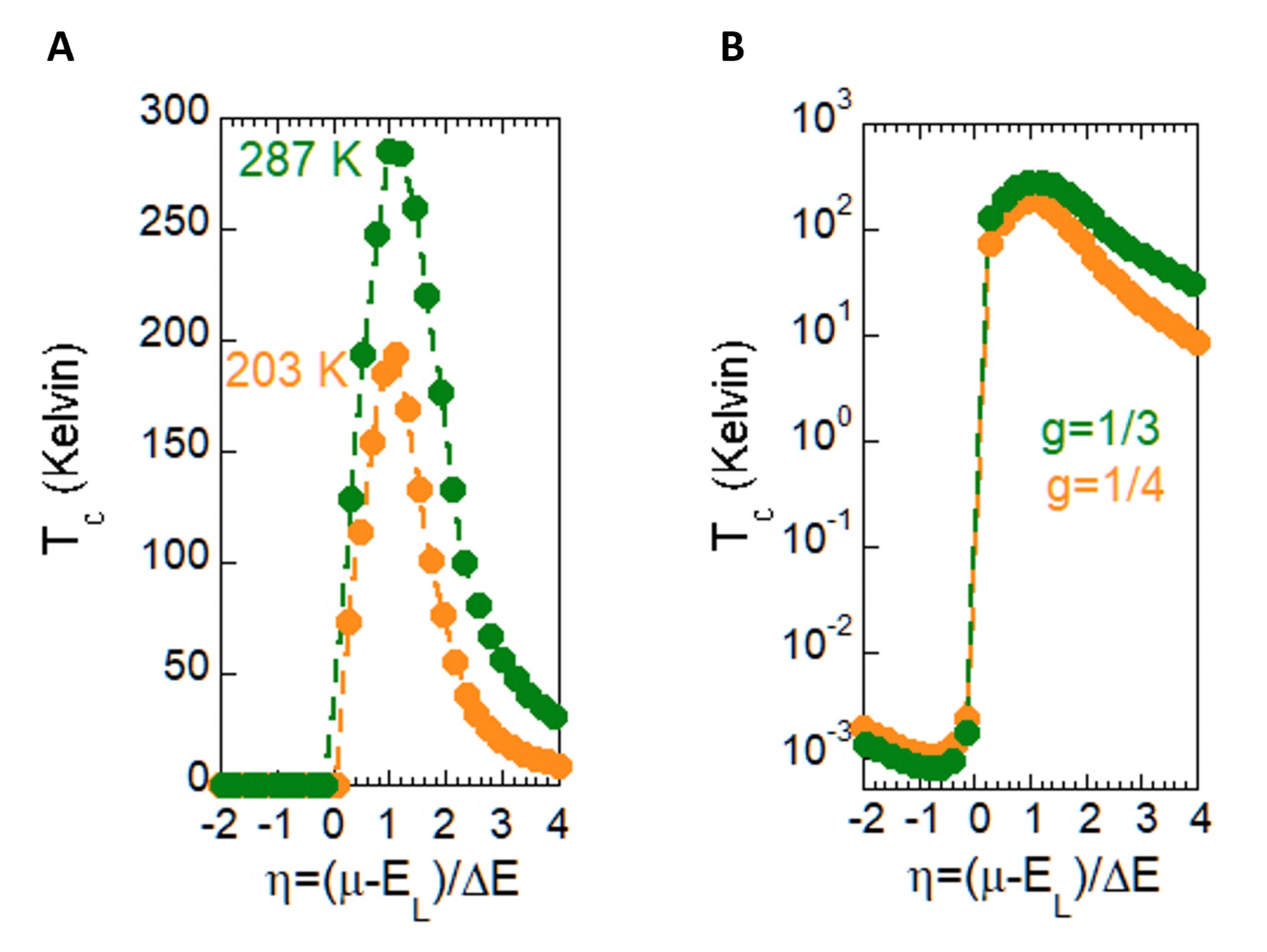}
\caption{The curves of  $T_C(\eta)$ versus $\eta$ at constant coupling in the upper subband $g= 0.33$ and $g= 0.25$ obtained by cuts of the 3D $dome$ shown in Fig. \ref{fig:4.4} in the linear (Panel A) and log (Panel B) scale reaching the maximum $T_C$ of $CSH_x$  and  $H_3S$ respectively. The asymmetric Fano line-shape of the resonance is shown in the log plot by the antiresonance for $\eta\simeq-0.5$ at the Lifshitz transition for the appearing of the new Fermi surface.}
\label{fig:4.5}     
\end{figure}
We propose the material design of a heterostructure of quantum wires showing room temperature superconductivity 
which reproduces the actual van Hove singularity at the Lifshitz transition calculated by band structure calculations of $H_3S$.
We start from the results on the Fano resonance in the high $T_C$ superlattice of quantum wires proposed  by Mazziotti et al. \cite{mazziotti2017possible} to simulate organic superconductors \cite{pinto2020potassium,barba2018anisotropic}.
We look for the theoretical prediction of the room  temperature superconducting  \textit{dome} like in $CSH_x$. 
In the heterostructure of quantum wires the electrons along the $x$-direction 
are free, while along the $z$-direction they see a periodic potential $V(z)$  forming 
quantum $n$ subbands characterized by quantized values of the transverse moment and the band index.
The size of superconducting wires $L$ and of intercalated spacers $W$ form  a periodic potential 
with period  $d = W + L $ and potential amplitude $V$.
We consider a case of weakly interacting  wires with a period comparable
 to the superconducting correlation length, and  the Fermi wavelength with $L=8.50$ \AA, $W=5.50$ \AA, $V=4.16\ meV$, which gives a band dispersion in the transversal direction due to electron hopping between wires $\Delta E=142 \ meV$.
The superconducting non-dimensional coupling constant $g_{nn'}$ for the three-band system 
and has a matrix structure that depends on the band indices $ n $ and $ n '$.
We choose the effective mass in the well $m_w = 0.86$, equal to the effective mass in the free $x$-direction, while $m_b = 1.0$  on the barrier, 
Following Ref.\cite{mazziotti2017possible} we have selected the intraband coupling in the lower two subbands $g_{ii}= 0.1$ and a weak  interband coupling constant $g_{ij}=0.1$ 
The theoretical superconducting $dome$ at a Fano resonance is obtained by changing the coupling $g=g_{33}$ in the third subband, appearing at the Lifshitz transition in the range $0.1< g< 0.49$ and the softening of the renormalized cut-off energy, $\tilde{\omega_0}$, as a function of the increasing pairing constant $g$ according to Migdal relation \cite{migdal1958interaction} 
$$
\tilde{\omega}_0=\omega_0  \sqrt{1-2g},
$$
with  $\omega_0 =230\ meV$ so that $\tilde{\omega}_0 =\Delta E$  for $g=0.3$.
The above relation stems from the screening effect of the electrons in the phonon propagator which was derived for a Fermi gas, we use here  to qualitatively estimate the effect of the phonon softening frequency with increasing coupling constant approaching to the Lifshitz transition for the appearing of the third Fermi surface.

The calculated superconducting $dome$ is shown in Fig. \ref{fig:4.4} where the critical temperature is plotted as a function of the coupling $g$ in the third subband and  of the Lifshitz parameter $\eta$. The yellow (green) dashed cut of the dome in Fig. \ref{fig:4.4}  is obtained with the fixed coupling  $g_{33}= 1/4$ ($g_{33} = 1/3$) reaching the maximum $T_C$ =$203 \ K$ ($287.7 \ K$) as observed in $SH_3$  ($SC_yH_x$).
The $T_C$($\eta$) curves, cuts of the $dome$ in Fig. \ref{fig:4.4}, are shown in Fig. \ref{fig:4.5}. The plot of log$(T_C)$ shows that, increasing $g_{33}$, the maximum of $T_C$ increases in the range $0.5<\eta<1$, at the 2D-1D dimensional crossover, while the minimum of log($T_C$) 
  is around $\eta\simeq-0.5$ decreases in agreement with the trend of a Fano resonance profile in the antiresonance regime.

\section{$Conclusions$}
 This work shows that a nanoscale superlattice of stripes made by  strain or dopants \cite{100} could be a generic features in organics, perovskites, and pressurized hydrides like $H_3S$ and $ CSH_x $ where the chemical potential is close to a Lifshitz transition \cite{101,102}. We have shown that  room  temperatures superconductivity in the weak coupling regime is possible in multigap superconductors by optimization of electron-phonon coupling in hot spots and tuning the contact exchange interaction between condensates. This work open new perspectives to develop new quantum complex materials including  Majorana fermions  \cite{mazziotti2018majorana} which open new venues to quantum devices.
Experimental advance in yttrium hydrides superconductors \cite{purans} has shown notable departures of the superconducting properties from the conventional Migdal, Eliashberg and BCS theories in ref. \cite{troyan} which indicate the presence of an additional mechanism of superconductivity which is identified here as the contact exchange interaction in the Fano resonances.

\acknowledgments
We thank Gaetano Campi, Andrea Perali and Roberto Raimondi for helpful discussions, and we acknowledge Superstripes onlus for financial support.


\begin{thebibliography}{0}
\bibitem{bcs} 
\Name {Bardeen J., Cooper L.N., \and Schrieffer J.R.} 
\REVIEW{Physical Review} {108} {1957} {1175}.
\bibitem{migdal1958interaction}
\Name {Migdal, A.B.} 
% Interaction between electrons and lattice vibrations in a normal metal. 
\REVIEW{JETP} { 7}  {1958} {996} 
\bibitem{mcm} 
\Name {W.I. McMillan}
\REVIEW{Physical Review} {167} {1968} {331}
\bibitem{100}
\Name {Snider, E., Dasenbrock-Gammon, N., McBride, $et$ $al.$}
 % title={Room-temperature superconductivity in a carbonaceous sulfur hydride},
 \REVIEW{Nature} {586} {2020} {373}.
\bibitem{Sakanishi2018first} 
  %{First-Principles study on superconductivity of P-and Cl-doped H3S},
\Name{Nakanishi, A., Ishikawa, T.\and Shimizu, K.}
\REVIEW{Journal of the Physical Society of Japan} {87} {2018} {124711}.
\bibitem{durajski2018gradual}
%{Gradual reduction of the superconducting transition temperature of H3S by partial replacing sulfur with phosphorus},
\Name{Durajski, A.P, \and Szczsniak, R.}
\REVIEW{Physica C: Superconductivity and its Applications} {554} {2018} {38}.
\bibitem{kruglov2017refined}
 % title={Refined phase diagram of the HS system with high-T c superconductivity},
\Name{Kruglov, I., Akashi, R., Yoshikawa, $et$ $al.$}
\REVIEW{Physical Review B} {96} {2017}{220101}
\bibitem{duan} 
\Name {Duan D., Liu Y., Tian F,  $et$ $al.$}
%Pressure induced metallization of dense (H$_2$S)2H$_2$ with high-T$_C$ superconductivity, 
\REVIEW{Sci. Rep.}  {4}  {2014} {6968}
\bibitem{droz} 
\Name {Drozdov, A. P., $et$ $al$}
 %Conventional superconductivity at 203 Kelvin at high pressures in the sulfur hydride system. 
\REVIEW{Nature}  {525}  {2015} {73}   
\bibitem{goncharov2017stable}
\Name {Goncharov, A. F., Lobanov, S. S., Prakapenka, V. B., \and Greenberg, E.}
%Stable high-pressure phases in the HS system determined by chemically reacting hydrogen and sulfur. 
\REVIEW{Physical Review B} {95} {2017} {140101}
\bibitem{duan2017structure}
\Name {Duan, D., Liu, Y., Ma, Y., Shao, Z., Liu, B., \and Cui, T.}
 %Structure and superconductivity of hydrides at high pressures. 
\REVIEW{National Science Review} {4} {2017} {121}.

\bibitem{bykova2020structure}
 %{Structure and composition of CSH compounds up to 143 GPa},
\Name{Bykova, E. Bykov, M.  Chariton, S. $et$ $al$}
\REVIEW{arXiv preprint} {2020} {arXiv:2012.10528}

\bibitem{bia15a} 
\Name {Bianconi A. \and Jarlborg T.}
 %Superconductivity above the lowest Earth temperature in pressurized sulfur hydride, 
\REVIEW{EPL Euro Phys. Lett.} {112} {2015} {37001}.
\bibitem{jar16}
\Name {Jarlborg T. \and Bianconi A.}
%Breakdown of the Migdal approximation at Lifshitz transitions with giant zero-point motion in the $H_3S$ superconductor,
\REVIEW{Scientific Reports}{6}{2016}{24816}.

\bibitem{bianconi1998high}
\Name {Bianconi, A.}
\title{Patent EP0733271A1 High Tc superconductors made by metal heterostructures at the atomic limit}
\REVIEW{European Patent Bulletin}  {98/22} {1998-05-27} 

\bibitem{bianconi2001process}
\Name {Bianconi A.}
\title{Patent US6265019B1 Process of increasing the critical temperature Tc of a bulk 
superconductor by making metal heterostructures at the atomic limit}
\REVIEW{Patent Official Gazette of the United States Patent and Trademark Office (PTO)}  {1248(4)} {Jul 24, 2001} {p.3989}

\bibitem{bianconi2013shape}
Bianconi, A. (2013). 
\Name {Bianconi, A.}
%Shape resonances in superstripes.
\REVIEW{Nature Physics} {9} {2013} {536} 
\bibitem{bianconi1994instability}
\Name {Bianconi, A. \and Missori, M.}
%The instability of a 2D electron gas near the critical density for a Wigner polaron crystal giving the quantum state of cuprate superconductors. 
\REVIEW{Solid State Communications} {91} {1994} {287} 
\bibitem{bianconi1996determination}
  %title={Determination of the Local Lattice Distortions in the Cu O 2 Plane of L a 1.85 S r 0.15 Cu O 4},
\Name{Bianconi, A., Saini, N.L., Lanzara, A. $et$ $al$}
\REVIEW{Physical Review Letters} {76} {1996} {3412}.
% strain in cuprates
\bibitem{2}
\Name {Agrestini, S., Saini, N. L., Bianconi, G. \and Bianconi, A }
%The strain of CuO2 lattice: the second variable for the phase diagram of cuprate perovskites. 
\REVIEW{ J. Phys. A: Math. Gen.}  {36} {2003}  {9133} 
\bibitem{08}
% Feshbach resonance and mesoscopic phase separation near a quantum critical point in multiband FeAs-based superconductors
\Name {Caivano, R., Fratini, M., Poccia, N., $et$ $al$} 
\REVIEW{ Superconductor Science and Technology} {22} {2008}  {014004}
\bibitem{09}
%Intrinsic arrested nanoscale phase separation near a topological Lifshitz transition in strongly correlated two-band metals %
\Name {Bianconi, A.; Poccia, N.; Sboychakov, A.O.; $et$ $al$}
\REVIEW{ Supercond. Sci. Technol.}  {28} {2015} {024005}.
\bibitem{ash}
 \Name{Ashcroft, N. W.} 
 %Symmetry and higher superconductivity in the lower elements. 
  \Book{Symmetry and Heterogeneity in High Temperature Superconductors}
  \Editor{A. Bianconi}, 
 % {NATO Science Series II: Math., Phys. and Chem.}
  \Vol{214}
  \Publ{Springer, Dordrecht}
  \Year{2006}
  \Page{3}.
 %{$doi 10.1007/1-4020-3989-1-1$}
% synchrotron 
 \bibitem{campi2019}
\Name {Campi, G. \and Bianconi, A.}
%Evolution of complexity in out-of-equilibrium systems by time-resolved or space-resolved synchrotron radiation techniques.
\REVIEW{Condensed Matter} {4} {2019} {32}
\bibitem{fratini2010scale}
  %title={Scale-free structural organization of oxygen interstitials in La 2 CuO 4+ y},
  \Name{Fratini, M.  $et$ $al$}
 \REVIEW{Nature} {466} {2010} {841}.
\bibitem{campi2015inhomogeneity}
  %title={Inhomogeneity of charge-density-wave order and quenched disorder in a high-T c superconductor},
  \Name{Campi, G.  $et$ $al$}
\REVIEW{Nature} {525} {2015} {359}
\bibitem{poccia2011evolution}
 % title={Evolution and control of oxygen order in a cuprate superconductor},
\Name{Poccia, N. $et$ $al$}
\REVIEW{Nature materials} {10} {2011} {733}.
\bibitem{campi2017nanoscale}
 % title={Nanoscale correlated disorder in out-of-equilibrium myelin ultrastructure},
 \Name{Campi, G.,  $et$ $al$}
 \REVIEW{ACS nano} {12} {2017} {729}.
\bibitem{della1995atomic}
 % title={Atomic and electronic structure probed by X-ray absorption spectroscopy: full multiple scattering analysis with the G4XANES package},
\Name{Della Longa, S., Soldatov, A., Pompa, M. \and Bianconi, A.}
\REVIEW{Computational Materials Science} {4} {1995} {199}.
\bibitem{darapaneni2020simulated}
 % title={Simulated field-modulated x-ray absorption in titania},
\Name{Darapaneni, P., Meyer, A. M., Sereda M. $et$ $al$}
\REVIEW{The Journal of Chemical Physics} {153} {2020} {054110}.
\bibitem{3}
\Name {Di Castro, D., Bianconi, G., Colapietro, $et$ $al$}
%Evidence for the strain critical point in high Tc superconductors. 
\REVIEW{Eur. Phys. J. B} {18} {2000} {617}.
\bibitem{4}
\Name {Kusmartsev, F.V., Di Castro, D., Bianconi, G.\and Bianconi, A. }
  %Transformation of strings into an inhomogeneous phase of stripes and itinerant carriers. 
\REVIEW{Physics Letters A} {275} {2000} {118}.
\bibitem{5}
\Name {Bianconi, A., Bianconi, G., Caprara, S.  $et$ $al$} 
%The stripe critical point for cuprates
\REVIEW{Journal of Physics: Condensed Matter} {12} { 2000 } {10655}.
\bibitem{6}
\Name {Makarov, I. A., Gavrichkov, V. A., Shneyder, E. I., $et$ $al$} 
%Effect of CuO 2 Lattice Strain on the Electronic Structure and Properties of High-Tc Cuprate Family. 
\REVIEW{J Supercond Nov Magn} {32} {2019} {1927}.
\bibitem{bianconi2000strain}
\Name {Bianconi, A., Saini, N.L., Agrestini, S., $et$ $al$} 
%The strain quantum critical point for superstripes in the phase diagram of all cuprate perovskites.
\REVIEW{Int. J. Mod. Phys. B} {14} {2000} {3342} 
\bibitem{8}
\Name {Bianconi, A., Di Castro, D., Bianconi, G.,  $et$ $al$}
%Coexistence of stripes and superconductivity: Tc amplification in a superlattice of superconducting stripes.
\REVIEW{Physica C: Superconductivity} {341} {2000} {1719} 

\bibitem{1}
\Name {Yamamoto, A., Takeshita, N., Terakura, C. \and Tokura, Y.}
 %High pressure effects revisited for the cuprate superconductor family with highest critical temperature.
\REVIEW{Nature Communications}  {6} {2015} {8990} {https://doi.org/10.1038/ncomms9990 }
%vanadates
\bibitem{9}
\Name {Di Gioacchino, D., Marcelli, A., Puri, A.  $et$ $al$} 
%Metastability phenomena in VO2 thin films. 
\REVIEW{Condensed Matter} {2} {2017} {10}
\bibitem{10}
\Name {Bianconi, A.}
%Multiplet splitting of final-state configurations in x-ray-absorption spectrum of metal 
%VO2: Effect of core-hole-screening, electron correlation, and metal-insulator transition. 
\REVIEW{Phys. Rev. B}  {26} {1982} {2741}
% diborides
\bibitem{12}
\Name {Bauer, E., Paul, C., Berger, S. $et$ $al$}
%Majumdar, S., Michor, H., Giovannini, M., .and Bianconi, A. (). 
%Thermal conductivity of superconducting MgB2. 
\REVIEW{Journal of Physics: Condensed Matter}   {13}  {2001}   {L487}
\bibitem{campi2006study}
%Study of temperature dependent atomic correlations in MgB2
\Name {Campi, G., Cappelluti, E., Proffen, Th. et al.}
\REVIEW{Eur. Phys. J. B} {52} {2006} {15}.
\bibitem{agrestini2004substitution}
%Substitution of Sc for Mg in MgB2: Effects on transition temperature and Kohn anomaly
\Name {Agrestini, S., Metallo, C., Filippi, M., $et$ $al$}
\REVIEW{Physical Review B} {70} {2004} {134514}
\bibitem{simonelli2009isotope}
%{Isotope effect on the E 2 g phonon and mesoscopic phase separation near the electronic topological transition in $Mg 1x Al x B 2$}
 \Name{Simonelli, L., Palmisano, V., Fratini, M. $et$ $al.$}
 \REVIEW{Phys. Rev. B} {80} {2009} {014520}
% iron based 
\bibitem{13}
\Name {Ricci, A., Poccia, N., Ciasca, G., Fratini, M., \and Bianconi} 
%The microstrain-doping phase diagram of the iron pnictides: heterostructures at atomic limit. 
\REVIEW{J Supercond Nov Magn} {22} {2009} {589}
\bibitem{ricci2010structural}
%{Structural phase transition and superlattice misfit strain of R FeAsO (R= La, Pr, Nd, Sm)},
\Name {Ricci, A., Poccia, N., Joseph, B. $et$ $al$}
 \REVIEW{Physical Review B} {82} {2010} {144507} 
%A15
\bibitem{weger} 
\Name {Barak G., Goldberg I.B. \and Weger M.}
%Band structure for some Nb3X compounds,
 \REVIEW{J. Phys. Chem. Solids}  {36} {1975}  {847}.
\bibitem{kle78}
\Name {Klein B.M., Boyer L.L., Papaconstantopoulos D.A., \and Mattheiss L.F.}
 %Self-consistent APW electronic structure calculations for the A15 compounds V3X and NbX, X=Al,Ga,Si,Ge and Sn,
\REVIEW{Phys. Rev. B}  {18}  {1978}  {6411}.
\bibitem{pic79} 
\Name {Pickett W.E., Ho K.M., \and Cohen M.L.}
%Electronic properties of Nb3Ge and Nb3Al from self-consistent pseudopotential band structure and density of states,
\REVIEW{Phys. Rev. B}  {19} {1979} {1734}
\bibitem{jar79} 
\Name {Jarlborg T.}
%Self-consistent LMTO band calculations on A15 compounds,
\REVIEW{J. Phys. F Metal Phys.} {9} {1979 } {283} 
\bibitem{kes78}
\Name {van Kessel, A.T., Myron H.W., \and Mueller, F.M.}
% Electronic structure of Nb3Sn
\REVIEW {Phys. Rev. Lett.}  {41} {978} {181}
\bibitem{pap77} 
\Name {Klein ,B.M., Boyer L.L., \and Papaconstantopoulos D.A.}
%Superconducting properties of A15 compounds derived from band-structure results,
\REVIEW{Phys. Rev. Lett.} {42} {1977} {530 } 
\bibitem{arb} 
\Name {Arbman G. \and Jarlborg T.}
%Trend studies of A15 compounds by self-consistent band calculations, 
\REVIEW{ Solid State Commun.}  {26} {1978} {1159}
\bibitem{jmp} 
\Name {Jarlborg T., Manuel A.A. \and Peter M.}
%Experimental and theoretical determination of the Fermi surface of $V_3$Si,
\REVIEW{Phys. Rev. B}  {27} {1983}  {4210}
\bibitem{jjp83}  
\Name {Jarlborg T., Junod  A. \and Peter  M.}
%Electronic structure, superconductivity and spin fluctuations in the A15 compounds V3X, Nb3X, X=Ir,Pt,Au. 
\REVIEW{Phys. Rev. B} {27}  {1983} {1558}
% recent A15
\bibitem{01}
% Structural instability and superconductivity in A-15 compounds 
\Name {Testardi, L. R.}
\REVIEW{Reviews of Modern Physics} {47} {1975} {637}
\bibitem{02}
%  High pressure induced shifts in the superconducting transition temperature of sputtered films.,
\Name {Chu, C. W., Testardi, L. R., \and Schmidt, P. H.}
\REVIEW{ Solid State Communications} {23} {1977} {841}
\bibitem{03}
% The effect of non-hydrostatic strain on the superconducting properties of in-situ formed Cu-Nb 3 Sn filamentary composites.
\Name {Bevk, J., Sunder, W., Hellman, F., and Geballe, T.}
\REVIEW{ IEEE Transactions on Magnetics } {21} {1985} {768}
\bibitem{04}
% Polytypism, polymorphism, and superconductivity in TaSe2-xTex
\Name {Luo, H., Xie, W., Tao, J., $et$ $al.$ } 
\REVIEW{Proc. Natl. Acad. Sci. USA} {112} {2015} {E1174} 
\bibitem{05}
% Phase separation in the vicinity of Fermi surface hot spots
\Name {Jaouen, T., Hildebrand, B., Mottas, $et$ $al.$ } 
\REVIEW{Physical Review B} {100} {2019} {075152}
\bibitem{06}
% Measuring the electron?phonon interaction in two-dimensional superconductors with He-atom scattering.
\Name {Benedek, G., Manson, J. R., Miret-Artes, $et$ $al.$ }.
\REVIEW{Condensed Matter} {5} {2020} {79}
\bibitem{07}
% Electronic topological transition in $Nb_3Al$ under compression: An ab initio study
\Name {Rajagopalan, M.}
\REVIEW{ Physica B: Condensed Matter} {413} {2013} {1}
%H3S
\bibitem{papa} 
\Name {Papaconstantopoulos D., Klein B.M., Mehl M.J. \and Pickett, W.E.}
%Cubic H$_3$S around 200 GPa: an atomic hydrogen superconductor stabilized by sulfur, 
\REVIEW{Phys. Rev. B} {91} {2015} {184511} 
%doi:10.1103/physrevb.91.184511
\bibitem{bia15} 
\Name {Bianconi A. \and Jarlborg T.}
%Lifshitz transitions and zero-point lattice fluctuations in sulfur hydride showing near room temperature superconductivity, 
\REVIEW{Nov. Supercond. Mater.} {1} {2015} {37}  {doi 10.1515/nsm-2015-0006}
\bibitem{fesi} 
\Name {Jarlborg T.}
%Electronic structure and properties of pure and doped FeSi from ab initio local-density theory,
\REVIEW{Phys. Rev. B} {59}  {1999} {15002}.
\bibitem{bron}
\Name {Jarlborg T., Chudzinski P.,  \and Giamarchi T.} 
%Effects of thermal and spin fluctuations on the band structure of purple bronze Li2Mo12O34
\REVIEW{ Phys. Rev. B}  {85} {2012 } {235108}
\bibitem{cevib} 
\Name {Jarlborg T.}
%Role of thermal disorder for magnetism and the transition in cerium: Results from density-functional theory
\REVIEW{Phys. Rev. B}  {89}   {2014}  {184426} 
\bibitem{ce}  
\Name {Jarlborg, T.,  Moroni, E.G., \and Grimvall G.}
 % Transition in Ce from temperature-dependent band-structure calculations
\REVIEW{Phys. Rev. B} {55} {1997} {1288} 
\bibitem{perali1996}
%The gap amplification at a shape resonance in a superlattice of quantum stripes: A mechanism for high Tc. 
\Name {Perali, A., Bianconi, A., Lanzara, A., \and Saini, N. L.}
\REVIEW{Sol. State Commun.} {100} {1996} {181}
\bibitem{bianconi1997high}
\Name {Bianconi, A., Valletta, A., Perali, A., \and Saini, N. L.}
%High Tc superconductivity in a superlattice of quantum stripes. 
\REVIEW{Solid State Communications} {102} {1997} {369}
\bibitem{valletta1997electronic}
\Name {Valletta, A., Bianconi, A., Perali, A., \and Saini, N. L.}
%Electronic and superconducting properties of a superlattice of quantum stripes at the atomic limit. 
\REVIEW{ Z Phys B  Con Mat} {104} {1997} {707}.
\bibitem{bianconi1998superconductivity}
\Name {Bianconi, A., Valletta, A., Perali, A. \and Saini, N. L.}
%Superconductivity of a striped phase at the atomic limit. 
\REVIEW{Physica C: Superconductivity} {296} {1998} {269}
\bibitem{perali2012anomalous}
 % title={Anomalous isotope effect near a 2.5 Lifshitz transition in a multi-band multi-condensate superconductor made of a superlattice of stripes},
\Name {Perali, A., Innocenti, D., Valletta, A.  \and Bianconi, A.}
\REVIEW{Supercond. Sci. Technol.} {25} {2012} {124002}
 \bibitem{Wentzel}
 \Name {Wentzel,G.}
 %Zur Quantenoptik,
 \REVIEW{ZS f. Physik} {22} {1924} {93}
\bibitem{Feynman}
 \Name {Feynman, R. P.}
 %Space-time approach to non-relativistic quantum mechanics. 
\REVIEW{Rev. Mod. Phys.,} {20}  {1948} {367}
 \bibitem{Fano}
\Name {Fano, U.}
  % Effects of configuration interaction on intensities and phase shifts. 
\REVIEW{Physical Review} {124} {1961} {1866}
\bibitem{Feshbach}
\Name {Feshbach, H.} 
\REVIEW{Annals of Physics } {19} {1962} {287}
\bibitem{Salomon}
\Name {Zhang, J.,  Van Kempen, E.,  Bourdel, T. $et$ $al.$} 
\REVIEW{Physical Review A} {70} {2004} {030702}
\bibitem{103}
 \Name {Bussmann-Holder, A., Keller, H., Simon A. \and Bianconi A.}
  \REVIEW{Condensed Matter}   {4}  {2019}  {91}
\bibitem{mazziotti2017possible}
\Name{Mazziotti, M. V., Valletta, A., Campi, G., $et$ $al.$}
 %Possible Fano resonance for high-Tc multi-gap superconductivity in p-Terphenyl doped by K at the Lifshitz transition. 
\REVIEW{EPL Europhysics Letters} {118} {2017} {37003}
\bibitem{mazziotti2021multigaps}
\Name {Mazziotti, M. V., Valletta, A., Raimondi, R., \and Bianconi, A.} 
%Multigaps superconductivity at unconventional Lifshitz transition in a 3D Rashba heterostructure at atomic limit. 
\REVIEW{Phys. Rev. B} {103} {2021} {024523}
\bibitem{barba2018anisotropic}
  %{Anisotropic thermal expansion of p-Terphenyl: a self-assembled supramolecular array of poly-p-phenyl nanoribbons},
\Name{Barba, L., Chita G., Campi, G. $et$ $al$}
\REVIEW{Journal of Superconductivity and Novel Magnetism} {31} {2018} {703}
\bibitem{pinto2020potassium}
%{Potassium-Doped Para-Terphenyl: Structure, Electrical Transport Properties and Possible Signatures of a Superconducting Transition},
\Name{Pinto, N., Di Nicola,C., Trapananti, A.  $et$ $al$}
\REVIEW{Condensed Matter} {5} {2020} {78}
 \Name {Jarlborg, T., \and Bianconi, A.}
%Multiple electronic components and Lifshitz transitions by oxygen wires formation 
 \REVIEW{Condensed Matter} {4} {2019} {15}
\bibitem{101}
 \Name {Bianconi, A.}
\REVIEW{Journal of Superconductivity} {18} {2005} {625}
\bibitem{102}
\Name {Kagan, M.Y., \and Bianconi, A.}
 %Fermi-Bose mixtures and BCS-BEC crossover in high-Tc superconductors. 
\REVIEW{Condensed Matter} {4} {2019}  {51}
\bibitem{mazziotti2018majorana}
\Name {Mazziotti, M. V., Scopigno, N., Grilli,M., \and Caprara, S.}
 %Majorana Fermions in One-Dimensional Structures at LaAlO3/SrTiO3 Oxide Interfaces. 
\REVIEW{Condensed Matter}  {3} {2018} {37}
\bibitem{purans}
\Name {Purans, J., Menushenkov, A. P., Besedin, S. P., $et$$al.$} 
%Local electronic structure rearrangements and strong anharmonicity in YH 3 under pressures up to 180 GPa. 
\REVIEW{Nature Communications} {12} {2021} {1}

\bibitem{troyan}
\Name {Troyan, I. A., Semenok, D. V., Kvashnin, A. G. $et$ $al.$} 
%Anomalous High?Temperature Superconductivity in YH6. 
\REVIEW{Advanced Materials} {33} {2021} {2006832} 


\end{thebibliography}
\end{document}